\documentstyle[preprint,tighten,aps]{revtex}    
\begin{document}           %
\draft
\preprint{\vbox{\noindent
Submitted to Physics Letters B\hfill INFNFE-10-94\\
          \null\hfill  INFNCA-TH-94-20}}
\title{Where are the beryllium neutrinos?
      }
\author{
         S.~Degl'Innocenti$^{(1)}$,
         G.~Fiorentini$^{(1)}$,
         and M.~Lissia$^{(2)}$
       }
\address{
$^{(1)}$Dipartimento di Fisica dell'Universit\`a di Ferrara, I-44100 Ferrara \\
       and Istituto Nazionale di Fisica Nucleare, Sezione di Ferrara,
        I-44100 Ferrara, Italy. \\
$^{(2)}$Istituto Nazionale di Fisica Nucleare, Sezione di Cagliari,
        I-09127 Cagliari,\\
        and Dipartimento di Fisica dell'Universit\`a di Cagliari,
        I-09100 Cagliari, Italy.
        }
\date{\today}
\maketitle                 
\begin{abstract}
We show that present experiments imply that neutrinos are nonstandard at
the 87\% C.L., independently of solar or nuclear physics.
Moreover, if neutrinos are standard, the $^7$Be flux must be almost
zero. Even if we arbitrarily disregard one of the experiments,
the neutrino flux must still be less than half of the value predicted
by standard solar models.
\end{abstract}
\narrowtext

It is a widespread statement that the Solar Neutrino Problem (SNP) is already
at the level of $^7$Be neutrinos, and it does not concern anymore just
the rare, hard to predict $^8$B neutrinos.

In this letter we intend to show clearly what is the essence of the  problem,
by
relating directly the results of solar neutrino experiments to
the fluxes of the $^7$Be, $^8$B and CNO neutrinos
($\Phi_{\text{Be}}$, $\Phi_{\text{B}}$ and $\Phi_{\text{CNO}}$) under the
hypothesis that neutrinos are standard (i.e. no mass, no mixing, no magnetic
moment, ...).

We shall employ methods similar to the ones we introduced in Ref.~\cite{1},
which have also been considered by other authors~\cite{2,3,4,5,6,7},
and keep into account the most recent experimental
results~\cite{8,9,10,11}.
Specifically, we shall address the following questions:
(1) What can be concluded about $\Phi_{\text{Be}}$ from the present
    experimental data for standard neutrinos?
(2) At which confidence level (C.L.) we can claim that either
    neutrinos are nonstandard or experimental results are incorrect?
(3) If neutrinos are standard and experimental results are correct,
    what is the predicted range for $\Phi_{\text{Be}}$?
(4) Can future experiments aimed at the measurement of the beryllium flux
    distinguish standard neutrinos from the
    Mikheyev-Smirnov-Wolfenstein (MSW) solution?

We have to  qualify in some detail the expression ``experimental results  are
correct''. In the rest of the paper, we  discuss separately the case when we
assume all experimental results to be correct, and when one of them is
disregarded. We use the standard procedure of quadratically combining
systematical and statistical errors for estimating confidence limits.
Alternative procedures are not obviously superior, and give similar results
as long as one does not go in the tails of the distributions.
Thus the sentence ``correct experimental result'' includes the statements
that errors are correctly evaluated by experimentalists,
and correctly treated by us.

Our discussion is practically independent of solar models, both standard and
non-standard ones.
Our main (very reasonable) assumption about the Sun is that the
present neutrino flux can be derived from the present value of the solar
constant (stationary Sun).
We also use the (very weak) assumption that the ratio
$x\equiv \Phi_{pep}/\Phi_{pp+pep}$ is
the same as in the Standard Solar Model (SSM), i.e.
$x_{\text{SSM}}=2.38\times10^{-3}$.
In fact, we made solar models with input parameters that vary wildly
out of the range defining the SSM, and found that this ratio is practically
constant.
Similarly, we take $y= \Phi_{N}/\Phi_{\text{CN}}=y_{\text{SSM}}=0.54$, and we
neglect the $hep$ and fluorine neutrinos.

Consequently, for standard neutrinos, i.e. neutrinos whose differential
flux is conserved independently of their energy from production to detection,
there remain only four unknowns:
\begin{equation}
 \Phi_{\text{Be}}\, ,\quad
 \Phi_{\text{B}} \, ,\quad
 \Phi_{pp+pep} \quad\text{and}\quad
 \Phi_{\text{CN}}\, .
\end{equation}
These four unknowns are constrained by exactly four equations, if all
three experimental data are correct and the Sun is stationary~\cite{12}.

Firstly, we have the luminosity equation
\begin{equation}
\label{lumi}
    K = \sum_i \left(\frac{Q}{2} - \langle E \rangle_i\right)\,\Phi_i \quad ,
\end{equation}
where $K$ is the solar constant, $Q$ is the energy released in the fusion
reaction $4p+2e \to \alpha + 2\nu$,
and $\langle E \rangle_i$ is the average neutrino energy of the $i$th
flux $\Phi_i$. Neglecting the contribution
of boron neutrinos, which is  expected to be even smaller than  the error on
$K$ (of the order of 0.2\%), and substituting the numerical constants,
Eq.~(\ref{lumi}) becomes:
\begin{equation}
\label{luminu}
 65.531 = 0.980\times\Phi_{pp+pep} + 0.939\times\Phi_{\text{Be}}
 + 0.937\times\Phi_{\text{CN}} \, ,
\end{equation}
where all fluxes, here and in the following, are in units of
$10^9\text{cm}^{-2}\text{s}^{-1}$.
We use this equation to express
$\Phi_{pp+pep}$ as function of the other unknown fluxes.

Therefore, the gallium experiments measure the combination:
\begin{equation}
\label{Gaeq}
S^{\text{Ga}}  = ( 79.75
           +    2.43\cdot 10^3  \times \Phi_{\text{B}}
           +    6.14 \times \Phi_{\text{Be}}
           +    7.49  \times \Phi_{\text{CN}} )\text{ SNU}  \, ,
\end{equation}
where SNU denotes solar neutrino units.
The weighting factors are given by the cross sections for neutrino capture on
gallium nuclei. The uncertainties as a function of energy
of these cross sections give
an overall uncertainty of $\pm 3\%$~\cite{13}.
We use as gallium experimental data the average between GALLEX and SAGE data:
\begin{equation}
\label{Gaexp}
S^{\text{Ga}}_{\text{exp}} = (74.4 \pm 9) \text{ SNU} \, .
\end{equation}
Since this error is definitely larger than the one on the neutrino-capture
cross section, we ignore this latter uncertainty.

With the same notations, the chlorine experiment  measures  the combination:
\begin{equation}
\label{Cleq}
S^{\text{Cl}}  = (  0.247
           +    1.09\cdot 10^3  \times \Phi_{\text{B}}
           +    0.236 \times \Phi_{\text{Be}}
           +    0.396 \times \Phi_{\text{CN}} )\text{ SNU}  \, ,
\end{equation}
where the cross section for $^8$B neutrino capture on Cl
has been taken from Ref.~\cite{14}. This time there is
an overall uncertainty of $\pm 1\%$.
The corresponding experimental result is:
\begin{equation}
\label{Clexp}
S^{\text{Cl}}_{\text{exp}} = (2.32 \pm 0.23) \text{ SNU} \, .
\end{equation}
Again we note that the error on the neutrino-capture cross section
can be neglected.

Finally, the Kamiokande experiment measures the $^8$B neutrino flux:
\begin{equation}
\label{Bexp}
\Phi_{\text{B}} = (2.9 \pm 0.42)\times 10^{-3}
\end{equation}
We can substitute the experimental data, Eqs.~(\ref{Gaexp}), (\ref{Clexp})
and (\ref{Bexp}), into
the two equations (\ref{Gaeq}) and (\ref{Cleq}), and solve them:
\begin{equation}
\Phi_{\text{Be}} = 4.9 \pm 3.4 \quad\text{ and }\quad
\Phi_{\text{CN}} = -5.7 \pm 5.6  \, .
\end{equation}
On the other hand, fluxes cannot be negative, so we stick to the most
favourable case $\Phi_{\text{CN}} = 0$, i.e. the one that
allows the highest $^7$Be flux.

Figure~\ref{fig1} shows the areas allowed at $1\sigma$ in the
(\/$^8$B, $^7$Be) plane
for $\Phi_{\text{CN}} = 0$.
We note that if all experiments are correct, the allowed areas intersect
in the unphysical region ($\Phi_{\text{Be}} \leq 0$).
Even if we disregard one of the three experimental results the
physical region,  $\Phi_{\text{Be}} > 0$,
looks disfavored.
At this point, we must point out that we are just demanding
the fluxes to be positive to say that the region is physical.
The introduction of any additional
physical information will further constrain the allowed region, e.g.
the knowledge about the nuclear reaction producing the $^8$B neutrinos
limits the values of the $^8$B  flux given the $^7$Be flux.

Thus we conclude that it appears unlikely that neutrinos  are
standard {\em and} at least two out of three experimental results
are  correct.

In an attempt to quantify this assertion, we make a chi-square
analysis. We take $\Phi_{\text{B}}$ and $\Phi_{\text{Be}}$
as free parameters, whereas we assume  $\Phi_{\text{CN}}=0$,
the case most favorable for standard neutrinos.
We evaluate the iso-$\chi^2$ curves both for the case when all experiments
are taken into account, and for the cases when one of them is disregarded
in turn (see Fig.~\ref{fig2}).
Statistics tells us, under the standard hypothesis, the probability that the
fluxes be in the region defined by $\chi^2< \chi^2_{\text{max}}$.
By requiring $\Phi_{\text{Be}} \ge 0$, we find that, when all experiments
are taken into account, there is
{\em at least} a 87\% probability that neutrinos are non standard.
Again we say
{\em at least}, since these are just the C.L. at which we can exclude
standard neutrinos at once without using any information about the
nuclear physics or about the Sun.
The analogous probabilities for the cases when one of the experiments is
disregarded are reported in Table~\ref{table1}. The most severe constraint
is given, as expected, by the combination of the chlorine and Kamiokande
data, that excludes standard neutrinos at the 88\% C.L.
On the contrary, the less severe
constraint is given, as already noticed in Refs.~\cite{12,15},
by the combination of the chlorine and gallium data: we can exclude
standard neutrinos without further hypothesis only at the 45\% C.L.

Let us take a somehow different attitude, and make some
model independent predictions testable in future experiments.
Assuming  that experimental results are correct and neutrinos are standard, we
determine the regions of the (\/$^8$B, $^7$Be) plane
that contains the true values of $\Phi_{\text{B}}$ and $\Phi_{\text{Be}}$
at the 90\%, 95\% and 98\% C.L.
These regions are also shown in Fig.~\ref{fig2}, and the corresponding
upper limits on the $\Phi_{\text{Be}}$ are reported in Table~\ref{table1}.
The combination of all three experiments
limit the $\Phi_{\text{Be}}$ to 0.6 at the
95\% C.L., and to 1.1 at the 98\% C.L. Even if we consider the least
restrictive case, and exclude the Kamiokande result, the $\Phi_{\text{Be}}$
must still be less than 2.1 at the 95\% C.L., and less than 2.7
at the 98\% C.L.

At this point, it is mandatory to recall that these upper values we
found for $\Phi_{\text{Be}}$ are very much lower than the values
predicted by solar models, both standard and non standard.
For reference, in our solar model we find that the beryllium flux
is $(4.79 \pm 0.24) \times 10^9\text{cm}^{-2}\text{sec}^{-1}$, and the
quoted upper values  are only 13\% and  45\% of this value at the 95\%
C.L., respectively for all experiments and for the {\em best} case, the one
that excludes Kamiokande.
Up to our knowledge, it is not possible construct solar models with such
a low $^7$Be flux without choices of input parameters and/or nuclear cross
sections that cannot be believed (a part for the case of an hypothetical
low energy $^{3}He +^{3}He$ resonance, see ~\cite{1}).
For the sake of discussion, we can use the approximate power-law
dependence of the fluxes on the central temperature $T_c$~\cite{12},
$\Phi_{\text{Be}}\sim T_c^{10}$ and $\Phi_{\text{B}}\sim T_c^{20}$,
and find that we need a 7\% central temperature reduction to reduce
the beryllium flux to half of its standard value. Correspondingly, the
boron flux is reduced to about $1/5$ of its standard value. These reduced
values barely fall inside the 98\% C.L. curve for the combination of
the chlorine and gallium data (such a $^8$B low flux is clearly
incompatible with Kamiokande).
However, we can produce solar models with such a large central
temperature reduction only at the price of large and unreasonable changes of
the solar parameters~\cite{12}.

If we combine this analysis  in the context of standard neutrinos
with corresponding analyses in the context of the MSW solution of
the SNP, which also predicts almost independently of the SSM a strong
reduction of the $^7$Be flux~\cite{14}, we have nowadays a very robust
prediction that
the future experiments aimed at the detection of $^7$Be neutrinos should
measure a $^7$Be flux considerably lower than the one predicted by SSM's.

It will be extremely interesting to see this low flux confirmed by future
experiments. The discrimination between different mechanisms that achieve such
low fluxes will, however, necessitate results from other experiments.

In conclusion,the combination of all present experiments tell us that
neutrinos are nonstandard at the 87\% C.L., even if we knew nothing about
the solar reactions that power our Sun. If we insist that neutrinos
are standard, the $^7$Be flux must, however, be less than 13\% of the
SSM value at the 95\% C.L.

If we are willing of throwing away one of the experimental data (no reason
to do that), the highest value of the $^7$Be flux compatible with standard
neutrino and two experimental data is
$2.7 \times 10^9\text{cm}^{-2}\text{sec}^{-1}$ at the 98\%, corresponding
to the combination of the chlorine and gallium data.

\begin{figure}
\caption[lmts1ex]{
We present in the (\/$^8$B, $^7$Be) plane
the regions allowed by the present experimental results.
Dashed lines correspond to
central values, solid lines denote $1\sigma$ limits.
We use the following experimental data:
($2.32\pm 0.23$)~SNU, ($74.4\pm 9.0$)~SNU, and
($2.9\pm 0.4$)$\times 10^6\text{cm}^{-2}\text{s}{-1}$ for the
chlorine (Cl), gallium (Ga), and Kamiokande (Ka) results,
respectively. We show the least stringent limits (see text), i.e.
when we assume a null CNO neutrino flux.
The diamond shows the fluxes predicted by our SSM with $1\sigma$ error bars.
               }
\label{fig1}
\end{figure}
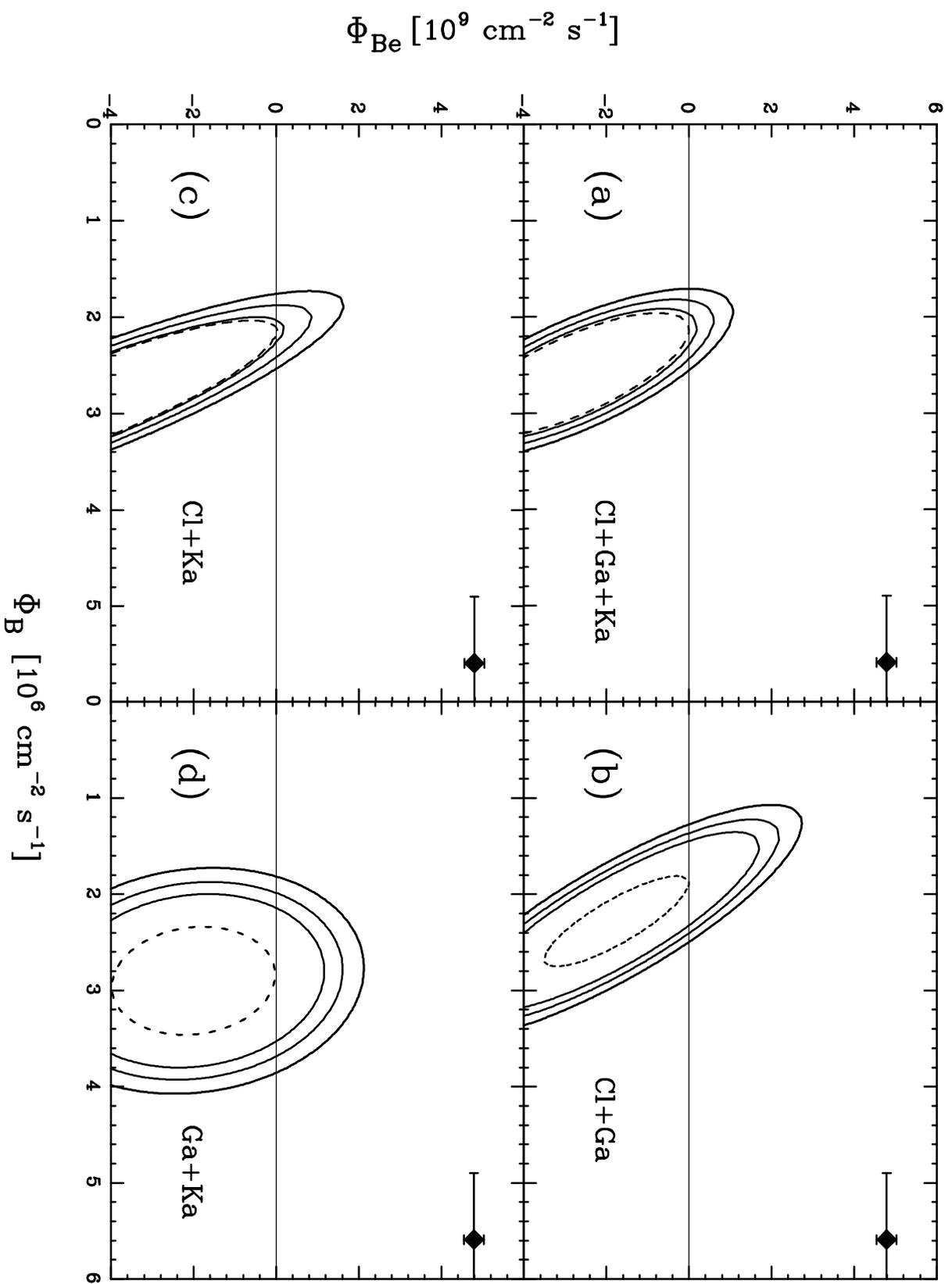
\begin{figure}
\caption[lmts23ex]{
  C.L. curves in the (\/$^8$B, $^7$Be) plane for: (a) chlorine + gallium +
  Kamiokande data, (b) chlorine + gallium data, (c) chlorine + Kamiokande data
  and (d) gallium + Kamiokande data. Solid curves correspond to 90\%, 95\% and
  98\% C.L., while the dashed curve corresponds to the C.L. at which the
  $^7$Be is negative (see Table).
  The diamonds show the fluxes predicted by our SSM with $1\sigma$ error bars.
               }
\label{fig2}
\end{figure}
\begin{table}
\caption[prob]{
The second column shows at which C.L. four different combinations of
present experimental data imply a negative $^7$Be flux, i.e. nonstandard
neutrinos (see text). The combinations are (a) chlorine + gallium +
Kamiokande, (b) chlorine + gallium, (c) chlorine + Kamiokande
and (d) gallium + Kamiokande.
The other columns show the maximum beryllium flux allowed by
the same combinations of data at the 90\%, 95\% and 98\% C.L.
The data correspond to the ones shown in Fig.~\protect\ref{fig2}.
         \label{table1}
         }
\begin{tabular}{lcccc}
 & C.L.         & \multicolumn{3}{c}{Maximum $^7$Be flux
                           [$10^9 \text{cm}^{-2} \text{s}^{-1}$]} \\
        & $^7\text{Be}\leq 0$ & 90\% C.L. & 95\% C.L.  & 98\% C.L.\\
\tableline
(a) Cl+Ga+Ka &     87\%          & 0.2    & 0.6        &  1.1 \\
(b) Cl+Ga    &     45\%          & 1.7    & 2.1        &  2.7 \\
(c) Cl+Ka    &     88\%          & 0.1    & 0.9        &  1.6 \\
(d) Ga+Ka    &     59\%          & 1.1    & 1.6        &  2.1 \\
\end{tabular}
\end{table}
\end{document}